\documentclass[final,5p,times,twocolumn]{elsarticle}

\usepackage{graphicx}
\usepackage{amssymb}
\usepackage{lineno}

\journal{Nuclear Instruments and Methods A}

\begin{document}

\begin{frontmatter}

\title{A new CVD Diamond Mosaic-Detector for (n,$\alpha$) Cross-Section Measurements at the n\_TOF Experiment at CERN}

\author[1,2] {C.~Wei{\ss}}
\ead{christina.weiss@cern.ch}
\author[1] {E.~Griesmayer}
\author[2] {C.~Guerrero}
\author[3] {S.~Altstadt}
\author[4] {J.~Andrzejewski}
\author[5] {L.~Audouin}
\author[1] {G.~Badurek}
\author[6] {M.~Barbagallo}
\author[7] {V.~B\'{e}cares}
\author[8] {F.~Be\v{c}v\'{a}\v{r}}
% ...
\author[] {\\(The n\_TOF Collaboration (www.cern.ch/ntof))}

\address[1]{Atominstitut, Technische Universit\"{a}t Wien, Austria}%
\address[2]{European Organization for Nuclear Research (CERN), Geneva, Switzerland}%
\address[3]{Johann-Wolfgang-Goethe Universit\"{a}t, Frankfurt, Germany}%
\address[4]{Uniwersytet \L\'{o}dzki, Lodz, Poland}%
\address[5]{Centre National de la Recherche Scientifique/IN2P3 - IPN, Orsay, France}%
\address[6]{Istituto Nazionale di Fisica Nucleare, Bari, Italy}%
\address[7]{Centro de Investigaciones Energeticas Medioambientales y Tecnol\'{o}gicas (CIEMAT), Madrid, Spain}%
\address[8]{Charles University, Prague, Czech Republic}%
% ...
%

\begin{abstract}
At the n\_TOF experiment at CERN a dedicated single-crystal chemical vapor deposition (sCVD) Diamond Mosaic-Detector has been developed for (n,$\alpha$) cross-section measurements. The detector, characterized by an excellent time and energy resolution, consists of an array of 9 sCVD diamond diodes. The detector has been characterized and a cross-section measurement has been performed for the $^{59}$Ni(n,$\alpha$)$^{56}$Fe reaction in 2012. The characteristics of the detector, its performance and the promising preliminary results of the experiment are presented. 

\end{abstract}

\begin{keyword}
sCVD diamond detectors, (n,$\alpha$) cross-section measurements, energy resolution, n\_TOF
\sep
\vspace{0.1cm}
\\CIVIDEC Instrumentation GmbH, Vienna, Austria, sponsored this work.



\end{keyword}

\end{frontmatter}


\section{Introduction}
\subsection{The n\_TOF Facility at CERN}
\begin{figure}[hbt] 
	\centering 
	\includegraphics[width=0.83\columnwidth,keepaspectratio]{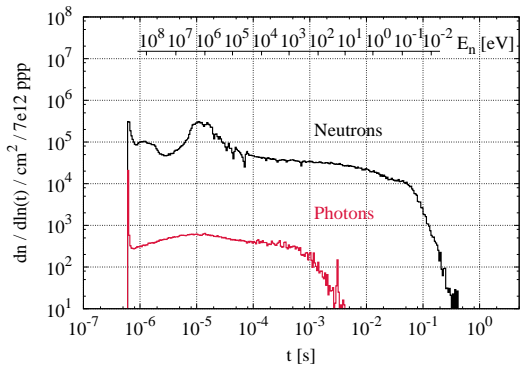}
	\caption{Arrival time of neutrons and photons at the measuring station at n\_TOF, with borated water as neutron moderator.}
	\label{fig:nTOF}
\end{figure}

The neutron time-of-flight facility n\_TOF at CERN \cite{nTOF} is devoted mainly to the measurement of neutron induced cross-sections. Neutrons are produced in a lead spallation target and moderated in a borated water-layer surrounding this target. The protons impinging on the target, provided by the CERN Proton Synchrotron (PS), have a momentum of 20 GeV/c and an intensity of 7$\cdot$10$^{12}$ protons/pulse. This provides a high instantaneous neutron fluence over a wide neutron energy (E$_n$) range, namely from thermal energies to GeV. \\
Background photons are produced during spallation (prompt $\gamma$ = $\gamma$-flash) as well as during neutron moderation (delayed $\gamma$ = in-beam $\gamma$-background) and arrive together with the neutron beam at the measuring station. In Figure \ref{fig:nTOF} the arrival time of photons and neutrons at the measuring station and the corresponding E$_n$ is shown. All charged particles produced during the spallation process are removed from the beam via a sweeping magnet in the beam line.\\
The measuring station is located 185 m downstream of the target. Due to the long flight path, E$_n$ can be determined via the time-of-flight technique with a very good energy resolution ($\Delta$E$_n$/E$_n$ = 3.2$\cdot$10$^{-4}$ for E$_n$=1 eV). \\
During the last 3 years, the experimental program at n\_TOF has been extended to neutron-induced charged-particle (n,x) reaction measurements, with focus on (n,$\alpha$) measurements. In this context, an innovative and very competitive detector has been developed for this purpose.

\subsection{The Task - (n,$\alpha$) Cross-Section Measurements}
\begin{figure}[hbt] 
	\centering 
	\includegraphics[width=0.45\columnwidth,keepaspectratio]{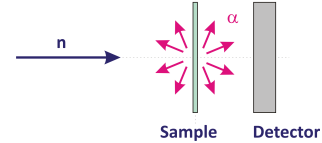}
	\caption{Schematic of the preferable experimental setup for (n,$\alpha$) cross-section measurements. The neutron beam, together with the in-beam $\gamma$-background, impinges on the sample, detector and the surrounding materials.}
	\label{fig:na}
\end{figure}

The experimental challenges for (n,$\alpha$) cross-section measurements are the following:
\begin{enumerate}
	\item (n,$\alpha$) reactions often feature small cross-sections.
	\item Thin samples are mandatory for low $\alpha$-absorption and low $\alpha$-energy loss in the sample.
	\item The measurements should be performed under vacuum to minimize the $\alpha$-energy loss on the way to the detector.
\end{enumerate}
To maximize the angular coverage and the yield, the detector is to be placed in the neutron beam in close proximity to the sample, see Figure \ref{fig:na}, and the size of the detector should be matched to the sample or the neutron beam-size respectively.
With such an experimental setup the background to be considered is caused by the in-beam $\gamma$-background as well as neutron-induced reactions in the detector and structural materials. This background may become significant and thus needs to be determined carefully during dedicated measurements. Therefore the samples have to be easily exchangeable, which calls for a compact detector-design.\\
To obtain optimal background rejection the following is required:
\begin{enumerate}
	\item High resolution $\alpha$-spectroscopy, so that different reactions can be identified according to the corresponding Q-value. This can be achieved by using a solid-state detector with low capacitance and a detector material which does not induce noise at room temperature.
	\item Low-Z material: This ensures a low interaction probability of $\gamma$-rays with the detector. In addition, (n,x) reactions in the detector material start at higher E$_n$.
	\item Minimized material budget: The thickness of the detector has to be adjusted to the range of the $\alpha$-particles in the detector. The backing layer should be as thin as possible. 
\end{enumerate}
To ensure these criteria were fulfilled, single-crystal chemical vapor deposition (sCVD) diamond (Z = 6) was chosen as detector material. 

\section{The Diamond Mosaic-Detector}
\begin{figure}[hbt] 
	\centering 
	\includegraphics[width=\columnwidth,keepaspectratio]{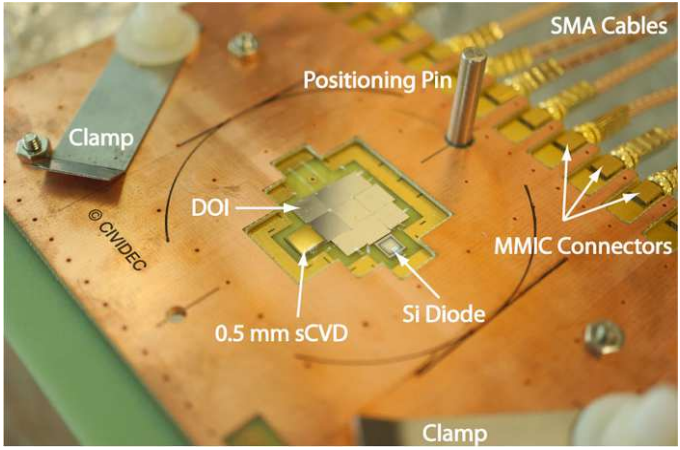}
	\caption{The Diamond Mosaic-Detector.}
	\label{fig:detector}
\end{figure}
A close-up picture of the Diamond Mosaic-Detector can be seen in Figure \ref{fig:detector}. For the cross-section measurement, the sample, a thin foil with a deposit of the isotope of interest, is positioned via the pin above the mosaic and pressed against the printed circuit board (PCB) with metallic clamps.\\
The mosaic consist of 8 sCVD diamond diodes and one diamond-on-Iridium (DOI) diode \cite{DOI1,DOI2}, each 150 $\mu$m in thickness and approximately 4 mm x 4 mm in size. The thickness of the diodes was chosen for spectroscopic measurements of $\alpha$ particles up to 20 MeV energy. \\
The diodes are metallized with 200 nm Al. On the front side of the diodes (seen in the picture) the metallization was left up to the edges of the diodes. On the reverse side of each diode an electrode of 3.8 mm x 3.8 mm was produced using negative photo-lithography.\\
An additional 500 $\mu$m thick sCVD diamond, metallized with Ti-Pt-Au, and a 100 $\mu$m thick Si diode can be seen in the picture. These two diodes were used for background measurements, which are not covered in this paper. \\
The diodes are glued and wire-bonded on a 0.5 mm gold-plated PCB and interconnected on the grounded front side of the detector. The high voltage (HV = 1V/$\mu$m) is individually applied to each diode from the read-out (back) side. 50 $\Omega$ MMIC connectors terminate the read-out lines. The capacitance of the mosaic channels is of the order of 10 pF. However, during measurements under vacuum, the capacitance was of the order of 35 pF, due to 20 cm long SMA cables and connectors inside the vacuum chamber.\\
At the rear of the detector-PCB an additional 0.5 mm copper-plated PCB is positioned for shielding purposes. At the front of the detector-PCB, a 1 mm copper-plated PCB with an opening for the diodes is positioned as a spacer between the sample and the diodes. The neutron beam at n\_TOF does not exceed the size of this opening, leaving the diodes and 1 mm of metallized PCB material in the beam. The thickness and design of the PCB plates were chosen as a reasonable compromise considering a minimum capacitance for each channel, a 50 $\Omega$ read-out line and a minimum material budget.\\
The PCB sandwich-structure and the full metallization on the grounded front side of the diodes provides shielding up to 2 GHz RF frequency.

\section{Dedicated Electronics}
\begin{figure}[hbt] 
	\centering 
	\includegraphics[width=0.83\columnwidth,keepaspectratio]{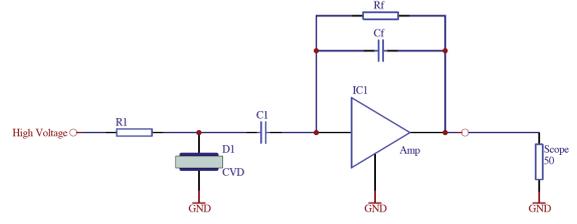}
	\caption{Circuit diagram with the CIVIDEC Cx shaping amplifiers.}
	\label{fig:Circuit}
\end{figure}
\begin{figure}[hbt] 
	\centering 
	\includegraphics[width=0.8\columnwidth,keepaspectratio]{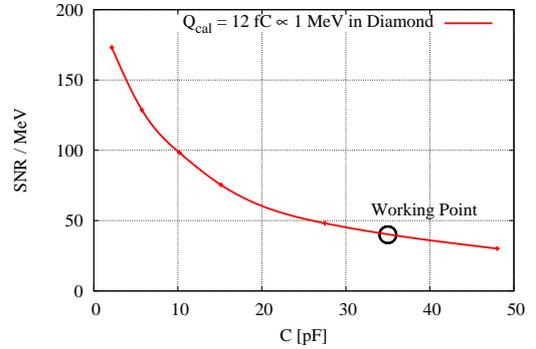}
	\caption{CIVIDEC Cx signal-to-noise ratio per MeV in diamond.}
	\label{fig:SNR}
\end{figure}

The CIVIDEC Cx spectroscopic shaping amplifiers \cite{CIVIDEC} were used for the measurements. The circuit diagram for the measurement is illustrated in Figure \ref{fig:Circuit}. The Cx amplifiers provide a Gaussian pulse shape with 180 ns shaping time and 80 ns rise time. The gain of the amplifiers was set to 8.2 mV/fC, being optimized for a total energy deposition in diamond of 20 MeV. The signal-to-noise ratio per MeV (SNR / MeV) as a function of the input capacitance of the amplifiers is illustrated in Figure \ref{fig:SNR}. For 35 pF one finds a SNR / MeV of 40 (corresponding to the experimental setup). The energy resolution for the setup is limited to FWHM = 50 keV and the corresponding time resolution is 2 ns / MeV, considering the rise time of the amplifiers.

\section{Energy Calibration}
The detector was calibrated with a multiple $\alpha$-source ($^{148}$Gd, $^{239}$Pu, $^{241}$Am, $^{244}$Cm), placing the uncollimated source at 1 mm distance from the detector in vacuum (p $\leq$ 1$\cdot$10$^{-4}$ mbar). A FLUKA simulation \cite{Battistoni,Ferrari} was performed for this setup to obtain the expected energy-spectra.
\begin{figure}[hbt] 
	\centering 
	\includegraphics[width=0.8\columnwidth,keepaspectratio]{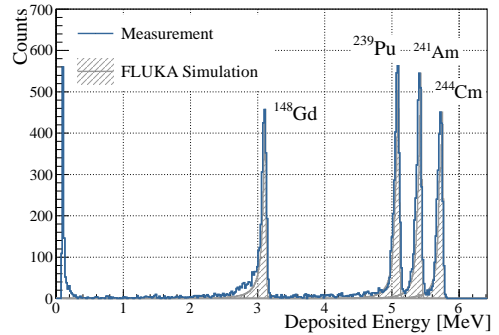}
	\caption{Calibration with multiple $\alpha$ source, exemplary for one sCVD diode, in comparison to simulations. The low energetic background (E $\leq$ 0.2 MeV) is related to $\gamma$ and e$^-$ emitted by the $\alpha$-source.}
	\label{fig:Calib}
\end{figure}
The recorded spectrum with one of the diodes in comparison with the FLUKA simulations is illustrated in Figure \ref{fig:Calib}. The energy resolution meets the expectations from the simulations with FWHM $<$ 70 keV at the $^{244}$Cm $\alpha$-peak, i.e. $\Delta$E/E $<$ 1.2\% at 5.8 MeV. 

\section{$^{59}$Ni(n,$\alpha$)$^{56}$Fe Cross-Section Measurement}
The Diamond Mosaic-Detector was used at n\_TOF in 2012 for a cross-section measurement of the $^{59}$Ni(n,$\alpha$)$^{56}$Fe reaction \cite{ND2013}. The sample consisted of a 25 $\mu$m Pt backing-foil on which 100 nm $^{59}$Ni (cross-section measurement) was electro-plated, followed by a 400 nm layer of $^6$LiF (neutron fluence measurement), with a diameter of 15 mm \cite{Harvey}. 
\begin{figure}[hbt] 
	\centering 
	\includegraphics[width=0.8\columnwidth,keepaspectratio]{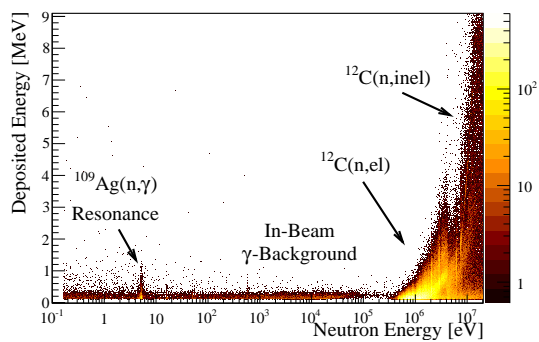}
	\caption{Background recorded at n\_TOF with the Diamond Mosaic-Detector, exemplary for one sCVD diode. The number of recorded signals is encoded in color.}
	\label{fig:SampleOut}
\end{figure}
Apart from the measurement of the sample, the following auxiliary measurements were performed:
\begin{enumerate}
	\item Background from the Diamond Mosaic-Detector, see Figure \ref{fig:SampleOut}. At low neutron energies, $\gamma$-rays from capture reactions in the surrounding materials, especially in $^{109}$Ag (glue for diodes), can be seen together with the in-beam $\gamma$-background, see also Figure \ref{fig:nTOF}. At about 200 keV, the elastic scattering of neutrons on $^{12}$C begins to be visible. At neutron energies above 6 MeV, the background is dominated by inelastic nuclear reactions on $^{12}$C.
	\item Background from the Pt foil.
	\item Background from the radioactive decay of $^{59}$Ni, which disintegrates with t$_{1/2}$ = (76 $\pm$ 5) kyr into $^{59}$Co via e$^-$-capture, producing predominantly 7 keV X-rays. No signals were recorded during this measurement, as expected for SNR / MeV = 40. 
\end{enumerate}
\begin{figure}[hbt] 
	\centering 
	\includegraphics[width=0.8\columnwidth,keepaspectratio]{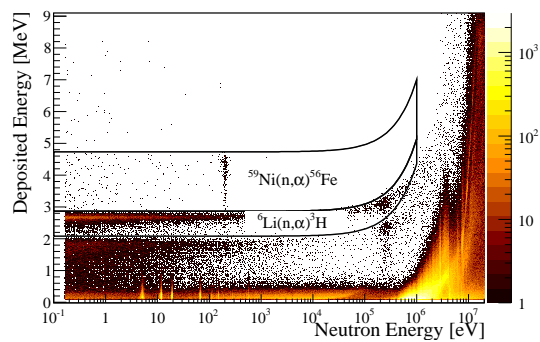}
	\caption{Experimental data recorded at n\_TOF with the $^{59}$Ni and $^6$Li sample in front of the Diamond Mosaic-Detector. The signals corresponding to the $\alpha$ particles of $^{59}$Ni(n,$\alpha$)$^{56}$Fe and the tritons from $^{6}$Li(n,$\alpha$)$^{3}$H are indicated. The number of recorded signals is encoded in color.}
	\label{fig:2D}
\end{figure}
The data recorded with one of the diodes during the measurement of the sample can be seen in Figure \ref{fig:2D}. The signals in the region labeled $^{59}$Ni(n,$\alpha$)$^{56}$Fe correspond to the $\alpha$ particles of this reaction. The dominant resonance at 203 eV is clearly visible. The region labeled $^{6}$Li(n,$\alpha$)$^{3}$H corresponds to signals created by the tritons of this reaction. The signals below this region correspond to the $\alpha$ particles of the $^{6}$Li(n,$\alpha$)$^{3}$H reaction, followed by the background signals from the detector and the Pt foil in the beam.\\
The cuts indicated in Figure \ref{fig:2D} are used for the selection of data for the cross-section and neutron fluence measurement respectively, and the detection efficiency corresponding to these cuts is calculated by fitting the peaks in the particle spectra (projection on the y-axis). The preliminary results for this measurement \cite{ND2013} are very promising and show that an accurate and clean (n,$\alpha$) cross-section measurement can be performed with this detection system.

\section{Conclusions}
The novel Diamond Mosaic-Detector has been developed for (n,$\alpha$) cross-section measurements at n\_TOF. The characteristics of the detector and the dedicated electronics were presented, together with results from the calibration with a $\alpha$-source and the $^{59}$Ni(n,$\alpha$)$^{56}$Fe cross-section measurement, which was performed at n\_TOF in 2012. \\ 
The sCVD diamond material has proven to be suitable for (n,$\alpha$) cross-section measurements in a heterogeneous beam, where background rejection is indispensable. The experimental data recorded with a $^{59}$Ni + $^6$Li sample in the n\_TOF neutron beam show a clear separation of the reaction products, see Figure \ref{fig:2D}.\\ 
The large-area, RF-tight Diamond Mosaic-Detector allowed a stable, spectroscopic measurement with an energy resolution $<$ 70 keV, in combination with the CIVIDEC Cx Shaping Amplifiers, see Figure \ref{fig:Calib}.

\section{Acknowledgements}
We are grateful to CIVIDEC Instrumentation GmbH for sponsorship. We acknowledge the contribution of C. Chery, R. De Oliveira, G. Ehrlich-Joop, D. Grenier, B. Hallgren, I. McGill, A. Mongelluzzo, H. Pernegger, M. Pomorski, P. Riedler and M. Schreck.



\begin{thebibliography}{00}
\bibitem{nTOF} C.~Guerrero et al., Eur. Phys. J. A, 49:27, (2013).
\bibitem{DOI1} M. Schreck et al., Appl. Phys. Lett. 78, 192 (2001).
\bibitem{DOI2} S. Gsell et al., Appl. Phys. Lett. 84, 4541 (2004).
\bibitem{CIVIDEC} www.cividec.at
\bibitem{Battistoni} G.~Battistoni et al., AIP Conference Proceeding 896, 31-49, (2007).
\bibitem{Ferrari} A.~Ferrari et al., CERN-2005-10 (2005), INFN/TC\_05/11, SLAC-R-773.
\bibitem{ND2013} C. Weiss et al., Nuclear Data Sheets, Proceedings ND2013.
\bibitem{Harvey} J.A. Harvey {\sc CONF-760715-P1} (1976).
\end{thebibliography}
\end{document}